\documentclass{emulateapj}
\usepackage{color}
\usepackage{ulem}

\shorttitle{The $r$-process in neutron star mergers}

\shortauthors{Wanajo et al.}

\begin{document}

\title{Production of all the \lowercase{$r$}-process nuclides in the
dynamical ejecta of neutron star mergers}

\author{Shinya Wanajo\altaffilmark{1}, 
        Yuichiro Sekiguchi\altaffilmark{2},
        Nobuya Nishimura\altaffilmark{3},
        Kenta Kiuchi\altaffilmark{2},
        Koutarou Kyutoku\altaffilmark{4},
        and 
        Masaru Shibata\altaffilmark{2}
        }

\altaffiltext{1}{iTHES Research Group, RIKEN, Wako, Saitama 351-0198, Japan;
        shinya.wanajo@riken.jp}

\altaffiltext{2}{Yukawa Institute for Theoretical Physics,
        Kyoto University, Kyoto 606-8502, Japan}

\altaffiltext{3}{Astrophysics, EPSAM, Keele University,
        Keele, ST5 5BG, UK}

\altaffiltext{4}{Department of Physics, University of
        Wisconsin-Milwaukee, P.O. Box 413, Milwaukee, WI 53201, USA}

\begin{abstract}
Recent studies suggest that binary neutron star (NS-NS) mergers robustly
produce the heavy $r$-process nuclei above the atomic mass number $A
\sim 130$ because of their ejecta consisting of almost pure neutrons
(electron fraction of $Y_\mathrm{e} < 0.1$). However, little production
of the lighter $r$-process nuclei ($A \approx 90$--120) conflicts with
the spectroscopic results of $r$-process-enhanced Galactic halo
stars. We present, for the first time, the result of nucleosynthesis
calculations based on the fully general-relativistic simulation of a
NS-NS merger with approximate neutrino transport. It is found that the
bulk of the dynamical ejecta are appreciably shock-heated and
neutrino-processed, resulting in a wide range of $Y_\mathrm{e}$
($\approx 0.09$--0.45). The mass-averaged abundance distribution of
calculated nucleosynthesis yields is in reasonable agreement with the
full-mass range ($A \approx 90$--240) of the solar $r$-process
curve. This implies, if our model is representative of such events, that
the dynamical ejecta of NS-NS mergers can be the origin of the Galactic
$r$-process nuclei. Our result also shows that the radioactive heating
after $\sim 1$~day from the merging, giving rise to $r$-process-powered
transient emission, is dominated by the $\beta$-decays of several
species close to stability with precisely measured half-lives. This
implies that the total radioactive heating rate for such an event can be
well constrained within about a factor of two if the ejected material
has a solar-like $r$-process pattern.
\end{abstract}

\keywords{
nuclear reactions, nucleosynthesis, abundances
--- stars: abundances
--- stars: neutron
}

\section{Introduction}\label{sec:intro}

The astrophysical site of the $r$-process, the rapid neutron-capture
process that makes half the elements heavier than iron, remains a
long-standing mystery of nucleosynthesis. Recently, compact binary
mergers (CBMs) of double neutron star (NS-NS) and black hole--neutron
star (BH-NS) systems have received considerable attention as possible
sources of the $r$-process nuclei \citep{Lattimer1974, Symbalisty1982,
Eichler1989, Meyer1989, Freiburghaus1999} according to the following
reasons.

First, radioactively powered ``kilonova'' emission from the
$r$-processed ejecta can be a promising electro-magnetic counterpart to
the gravitational-wave signal from a CBM event \citep{Li1998,
Metzger2010, Goriely2011, Kasen2013, Barnes2013, Tanaka2013,
Grossman2014}. The possible identification of a kilonova associated with
the \textit{Swift} GRB~130603B \citep{Berger2013, Tanvir2013} also
indicates that CBMs are the progenitors of short-duration gamma-ray
bursts and the sources of $r$-process elements \citep{Hotokezaka2013a,
Tanaka2014}.

Another reason is that core-collapse supernovae (CCSNe; in particular
proto-NS wind), the site that has been believed to be the promising
sources of the $r$-process nuclei, are found to provide only marginal
conditions for making the elements beyond iron \citep{Martinez2012,
Roberts2012, Fischer2012}. Nucleosynthesis studies with such physical
environments confirm that CCSNe produce the elements only up to the
atomic mass number $A \sim 110$ \citep{Wanajo2011, Wanajo2013}. One
possible exception could be the scenario of (still hypothetical) rapidly
rotating, strongly magnetized CCSN cores \citep{Winteler2012}.

Recently, \citet{Goriely2011} and \citet{Bauswein2013} have explored
nucleosynthesis based on the approximate (conformally flat spatial
metric) general-relativistic (GR) simulations of NS-NS mergers. They
found that the ejecta had extremely low electron fractions
($Y_\mathrm{e} < 0.1$), which led to fission recycling and thus robust
production of only heavy $r$-process nuclei with $A \gtrsim 130$.
Similar results were obtained from the Newtonian simulations of NS-NS
and BH-NS mergers by \citet{Roberts2011, Korobkin2012, Rosswog2014}.

Little production of the lighter $r$-process nuclei ($A \approx
90$--120) conflicts, however, with the recent spectroscopic results of
Galactic halo stars \citep{Sneden2008, Siqueira2014}. That is, the
so-called ``universality'' of the (solar-like) $r$-process pattern,
first identified for $Z \gtrsim 56$ ($A \gtrsim 140$), persists down to
$Z \sim 38$ ($A \sim 90$) within about a factor of two. There has been
no sign of nucleosynthetic events making the nuclei exclusively with $A
\gtrsim 130$. Contribution from, e.g., the subsequent BH accretion-torus
wind \citep{Surman2008, Wanajo2012, Fernandez2013} might cure this
problem.

In this Letter, we report our first result of nucleosynthesis study
based on the full-GR, approximate neutrino transport simulation of a
NS-NS merger. The GR effects, being crucial for the dynamical evolutions
of merger ejecta as pointed out by \citet{Hotokezaka2013b}, were not
fully taken into account in the previous studies. Moreover, neutrino
transport that can affect the ejecta $Y_\mathrm{e}$ is neglected in all
previous studies \citep[except for the 2D Newtonian simulation
by][without nucleosynthesis calculations]{Dessart2009}. Our NS-NS merger
model is described in Section~\ref{sec:model}. The subsequent
nucleosynthesis result is presented in
Section~\ref{sec:nucleosynthesis}. The radioactive heating rates
(relevant for kilonova emission) are also obtained from the
nucleosynthesis calculations (Section~\ref{sec:heating}).

\section{Merger Model}\label{sec:model}

\begin{figure}
\epsscale{1.0}
\plotone{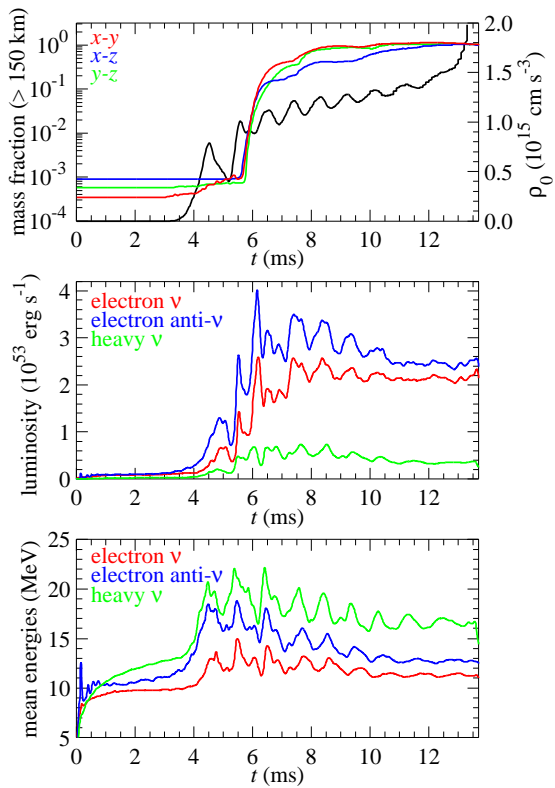}

\caption{Temporal evolutions of ejecta mass fractions outside 150~km
 from the origin of the coordinate axis for the $x$-$y$, $x$-$z$, and
 $y$-$z$ planes in the (2000~km)$^3$ cube (see Figure~\ref{fig:image};
 with the width $\approx 13$~km for each plane). The ejecta mass ratio
 at the end of simulation is $\sim 5 : 2 : 3$ for these planes.  The
 masses at $t = 0$ are due to the background medium, whose fractions are
 sufficiently small compared to the total masses.  Also shown is the
 temporal evolution of density at the origin.  The middle and bottom
 panels display, respectively, the luminosities and angle-averaged mean
 energies for $\nu_e$, $\bar{\nu}_e$, and heavy-lepton neutrinos. Note
 that the neutrinos of $\sim 10$~MeV at $t \lesssim 4$~ms are
 unimportant because of the low luminosities.}

\label{fig:evolution}
\end{figure}

\begin{figure*}
\includegraphics[width=0.45\textwidth, angle=-90]{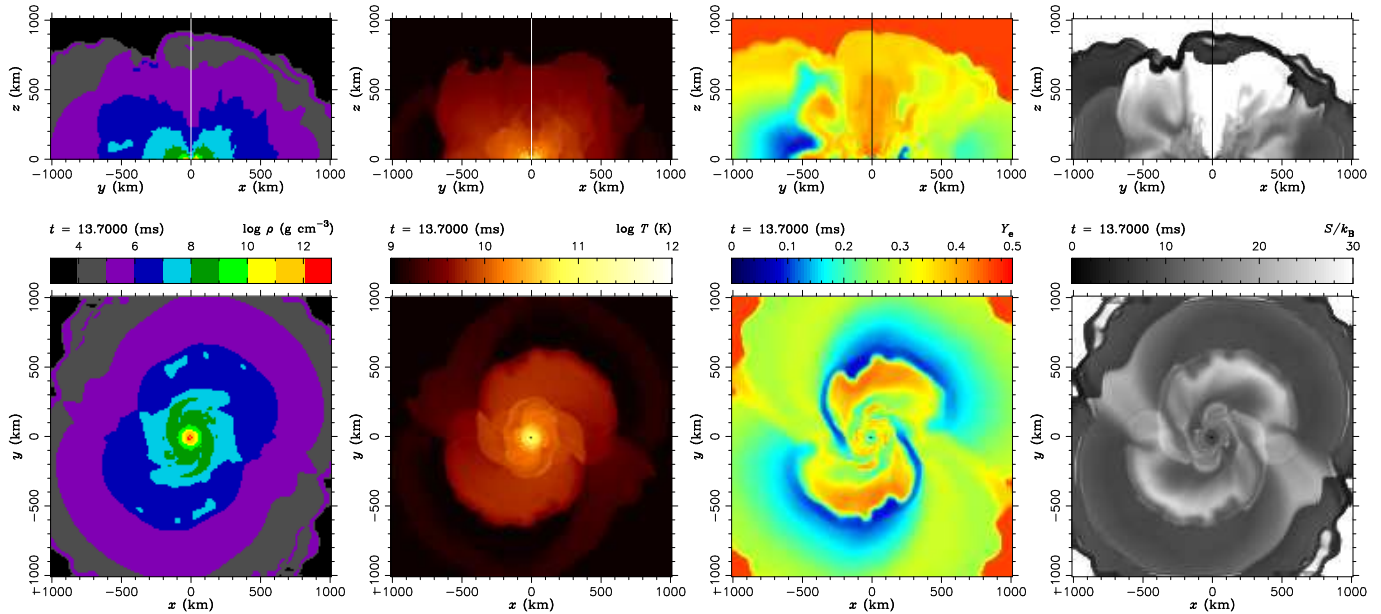}

\caption{Color-coded distributions for density, temperature,
 $Y_\mathrm{e}$, and $S/k_\mathrm{B}$ (from left to right) on the
 $x$-$y$ (lower panels), $x$-$z$ (positive sides of top panels), and
 $y$-$z$ (negative sides of top panels) planes at the end of
 simulation.}

\label{fig:image}
\end{figure*}

The hydrodynamical evolution of a NS-NS merger is followed with a
recently-developed 3D full-GR code (Y. Sekiguchi et al. 2014, in
preparation), which is updated from the previous version
\citep{Sekiguchi2010, Sekiguchi2011a, Sekiguchi2011b}. Neutrino
transport is taken into account based on the Thorne's moment scheme
\citep{Thorne1981, Shibata2011} with a closure relation. For neutrino
heating, absorption on free nucleons is considered. The gravitational
masses (in isolation) are taken to be $1.3\, M_\odot$ for both NSs.

We adopt an equation of state (EOS) of dense matter developed in
\citet[][SFHo]{Steiner2013}, which has the maximum NS mass sufficiently
greater than the largest well-measured mass \citep[$\approx 2\,
M_\odot$,][]{Demorest2010, Antoniadis2013}. This EOS gives the radius
$\approx 12$~km for a cold NS, being in the range constrained from
nuclear experiments, nuclear theory, and astrophysical observations,
10.7--13.1~km for a $1.4\, M_\odot$ NS \citep{Lattimer2013}. Note that
the EOS of \citet{Shen1998} adopted in many previous simulations gives
$\approx 14.5$~km for a $1.4\, M_\odot$ NS, being substantially greater
than the upper bound of this constraint.

At the beginning of simulation ($t = 0$), each NS consists of matter
with $Y_\mathrm{e} \approx 0.06$ in the (neutrino-less)
$\beta$-equilibrium with the constant temperature 0.1~MeV. The
background medium is placed with the same temperature, density
decreasing from $10^5$~g~cm$^{-3}$ (in the central region) to
$10^3$~g~cm$^{-3}$, and $Y_\mathrm{e} = 0.46$.  The merging of NSs
starts at $t \sim 3$~ms with increasing density at the origin of the
coordinate axis, $\rho_0$ (Figure~\ref{fig:evolution}). This leads to
the steep rises of masses ($t \sim 5.5$~ms) outside 150~km (from the
center) coming from the contact interface region.

A hypermassive NS (HMNS) forms at $t \sim 4.5$~ms. The second phase of
mass ejection follows in response to the interaction between the inner
atmospheric material (originating from the shear interface) and the
rapidly rotating, quasi-radially oscillating HMNS (from $t \sim
7.5$~ms). We find that the total ejecta mass is dominated ($\sim 60\%$)
by this second phase. The simulation ends at $t = 13.7$~ms with the
distributions of density, temperature, $Y_\mathrm{e}$, and entropy (per
nucleon; $S/k_\mathrm{B}$, $k_\mathrm{B}$ is Boltzmann's constant) shown
in Figure~\ref{fig:image}.\footnote{Movies of the simulation are
available from http://cosnucs.riken.jp/movie.html.}  At this time, the
bulk of ejecta (total mass of $M_\mathrm{ej} \approx 0.01\, M_\odot$)
are freely expanding with the velocities $\sim (0.1$--$0.3) c$ ($c$ is
the speed of light). 

The behavior of mass ejection described here is in qualitative agreement
with the previous full-GR \citep[][for soft EOSs]{Hotokezaka2013b} and
approximate GR \citep[][for the same SFHo EOS]{Bauswein2013} works. As
pointed out in these studies, the mass ejection is due to shock-heating
and tidal torque; neutrino-heating plays a subdominant role.

As the HMNS forms, temperature near its surface gets as high as $\sim
10$~MeV ($\sim 100$~GK), giving rise to copious $e^- e^+$ pairs that
activate the weak interactions $n + e^+ \rightarrow \bar{\nu}_e + p$, $p
+ e^- \rightarrow \nu_e + n$, and their inverses. The $e^+$ and $\nu_e$
captures convert some part of neutrons to protons; the ejecta
$Y_\mathrm{e}$'s increase from the initial low values.\footnote{The fast
moving NSs and subsequent merger ejecta in the background medium make
shocks that can increase temperature and thus $Y_\mathrm{e}$. However,
the mass suffering from these artifacts is negligibly small compared to
the total ejecta mass (Figure~\ref{fig:evolution}).}  The first outgoing
ejecta from the contact interface region are away from the HMNS when
it forms and thus neutrino capture is subdominant. As a result, the
$Y_\mathrm{e}$'s are relatively low ($\sim 0.1$--0.2; blue-cyan spiral
arms in Figure~\ref{fig:image}). The outer ejecta with higher
$Y_\mathrm{e}$ ($\sim 0.2$--0.3) are unimportant in the total ejecta
mass because of their low densities.

In the second phase of mass ejection, neutrinos coming from the HMNS
surface play a crucial role. The luminosities and mean energies are only
slightly greater for $\bar{\nu}_e$ than those for $\nu_e$
(Figure~\ref{fig:evolution}). The asymptotic $Y_\mathrm{e}$ (after
sufficient time) with these values is expected to be $Y_\mathrm{e, a}
\sim 0.5$ \citep[e.g., eq.~(77) in][]{Qian1996}. However, neutrino
absorption in the fast outgoing ejecta freezes before $Y_\mathrm{e}$
reaches $Y_\mathrm{e, a}$, resulting in $Y_\mathrm{e} \sim 0.3$--0.4
(yellow-orange spiral arms in Figure~\ref{fig:image}).

The ejecta mass distributions in $Y_\mathrm{e}$ and $S/k_\mathrm{B}$ at
the end of simulation are displayed in Figure~\ref{fig:histogram} for
the $x$-$y$, $x$-$z$, and $y$-$z$ planes. We find that the
$Y_\mathrm{e}$'s widely distribute between 0.09 and 0.45 with greater
amounts for higher $Y_\mathrm{e}$, in which the initial
$\beta$-equilibrium values ($\approx 0.06$) have gone. Non-orbital
ejecta have higher $Y_\mathrm{e}$'s because of the shock-heated matter
escaping to the low-density polar regions \citep{Hotokezaka2013b}. The
shock heating results in $S/k_\mathrm{B}$ up to $\approx 26$ and 50 for
the orbital and non-orbital planes, respectively (with higher values for
higher $Y_\mathrm{e}$), which are sizably greater than those in
\citet[][$S/k_\mathrm{B} \sim 1$--3]{Goriely2011} with the Shen's EOS.

\section{The \lowercase{$r$}-process}\label{sec:nucleosynthesis}

\begin{figure}
\epsscale{1.0}
\plotone{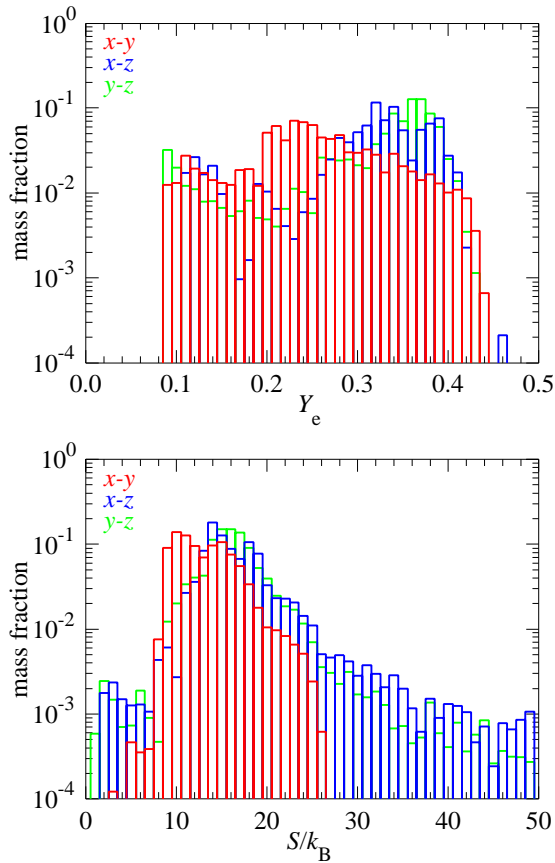}

\caption{Mass fractions outside 150~km from the center
 vs. $Y_\mathrm{e}$ (top) and $S/k_\mathrm{B}$ (bottom) at the end of
 simulation for the $x$-$y$, $x$-$z$, and $y$-$z$ planes. The widths of
 $Y_\mathrm{e}$ and $S/k_\mathrm{B}$ are chosen to be $\Delta
 Y_\mathrm{e} = 0.01$ and $\Delta S/k_\mathrm{B} = 1$,
 respectively.}

\label{fig:histogram}
\end{figure}

The nucleosynthesis analysis makes use of the thermodynamic trajectories
of the ejecta particles traced on the orbital plane. A representative
particle is chosen from each $Y_\mathrm{e}$-bin (from $Y_\mathrm{e} =
0.09$ to 0.44 with the interval of $\Delta Y_\mathrm{e} = 0.01$
(Figure~\ref{fig:histogram}). For simplicity, we analyze only the
$x$-$y$ components because of the dominance of the ejecta masses close
to the orbital plane. Each nucleosynthesis calculation is initiated when
the temperature decreases to 10~GK, where the initial composition is
given by $Y_\mathrm{e}$ and $1 - Y_\mathrm{e}$ for the mass fractions of
free protons and neutrons.

The reaction network consists of 6300 species from single neutrons and
protons to the $Z = 110$ isotopes. Experimental rates, when available,
are taken from the latest versions of
REACLIB\footnote{https://groups.nscl.msu.edu/jina/reaclib/db/index.php.}
\citep{Cyburt2010} and Nuclear Wallet Cards
\footnote{http://www.nndc.bnl.gov/wallet/}. Otherwise, the theoretical
estimates of fusion
rates\footnote{http://www.astro.ulb.ac.be/pmwiki/Brusslib/Brusslib.}
\citep[TALYS,][]{Goriely2008} and $\beta$-decay half-lives
\citep[GT2,][]{Tachibana1990} are adopted, where both are based on the
same nuclear masses \citep[HFB-21,][]{Goriely2010}. Theoretical fission
properties adopted are those estimated on the basis of the HFB-14 mass
model. For fission fragments, a Gaussian-type distribution is assumed
with emission of four prompt neutrons per event. Neutrino captures are
not included, which make only slight shifts of $Y_\mathrm{e}$ (typically
an increase of $\sim 0.01$ from 10~GK to 5~GK).

The hydrodynamical trajectories end with temperatures $\sim
5$~GK. Further temporal evolutions are followed by the density drop like
$t^{-3}$ and with the temperatures computed with the EOS of
\citet{Timmes2000} by adding the entropies generated by $\beta$-decay,
fission, and $\alpha$-decay. This entropy generation slows the
temperature drop around 1~GK \citep[e.g.,][]{Korobkin2012}. The effect
is, however, less dramatic than those found in previous works because of
the higher ejecta entropies in our result.

\begin{figure}
\epsscale{1.0}
\plotone{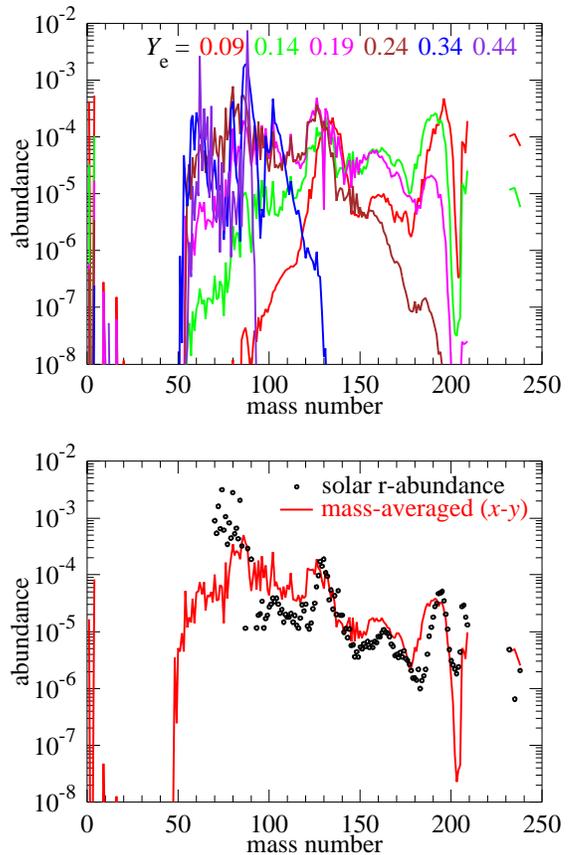}

\caption{Final nuclear abundances for selected trajectories (top) and
 that mass-averaged (bottom; compared with the solar $r$-process
 abundances).}

\label{fig:ymav}
\end{figure}

Figure~\ref{fig:ymav} (top) displays the final nuclear abundances for
selected trajectories. We find a variety of nucleosynthetic outcomes:
iron-peak and $A \sim 90$ abundances made in nuclear quasi-equilibrium
for $Y_\mathrm{e} \gtrsim 0.4$, light $r$-process abundances for
$Y_\mathrm{e} \sim 0.2$--0.4, and heavy $r$-process abundances for
$Y_\mathrm{e} \lesssim 0.2$. Different from the previous works, we find
no fission recycling; the nuclear flow for the lowest $Y_\mathrm{e}$ ($=
0.09$) trajectory reaches $A \sim 280$, the fissile point by
neutron-induced fission, only at the freezeout of
$r$-processing. Spontaneous fission plays a role for forming the $A \sim
130$ abundance peak, but only for $Y_\mathrm{e} < 0.15$.

Figure~\ref{fig:ymav} (bottom) shows the mass-averaged nuclear
abundances by weighting the final yields for the representative
trajectories with their $Y_\mathrm{e}$ mass fractions on the orbital
plane (Figure~\ref{fig:histogram}). We find a good agreement of our
result with the solar $r$-process abundance distribution over the
full-$A$ range of $\sim 90$--240 (although the pattern would be somewhat
modified by adding non-orbital components). This result, differing from
the previous works exhibiting production of $A \gtrsim 130$ nuclei only,
is a consequence of the wide $Y_\mathrm{e}$ distribution predicted from
our full-GR, neutrino transport simulation. Note also that fission plays
a subdominant role for the final nucleosynthetic abundances. The second
($A\sim 130$) and rare-earth-element ($A \sim 160$) peak abundances are
dominated by direct production from the trajectories of $Y_\mathrm{e}
\sim 0.2$. Our result reasonably reproduces the solar-like abundance
ratio between the second ($A \sim 130$) and third ($A \sim 195$) peaks
as well, which is difficult to explain by fission recycling.

Given that the model is representative of NS-NS mergers, our result
gives an important implication; the dynamical ejecta of NS-NS mergers
can be the dominant origin of all the Galactic $r$-process nuclei. Other
contributions from, e.g., the BH-torus wind after collapse of HMNSs, as
invoked in the previous studies to account for the (solar-like)
$r$-process universality, may not be needed. The amount of entirely
$r$-processed ejecta $M_\mathrm{ej} \approx 0.01\, M_\odot$ with present
estimates of the Galactic event rate \citep[a few $10^{-5}$~yr$^{-1}$,
e.g.,][]{Dominik2012} is also compatible with the mass of the Galactic
$r$-process abundances as also discussed in previous studies
\citep[][]{Korobkin2012, Bauswein2013}.

\section{Radioactive Heating}\label{sec:heating}

\begin{figure*}
\epsscale{1.0}
\plotone{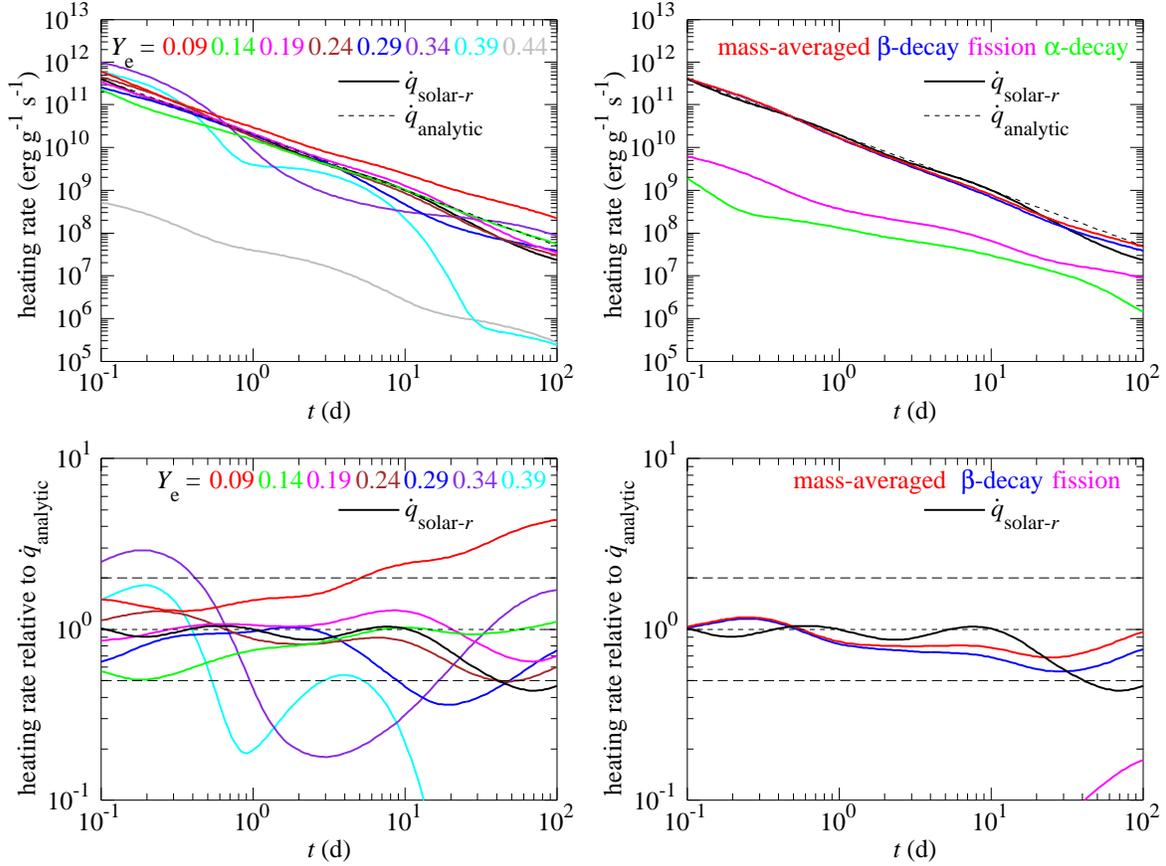}

\caption{Heating rates as functions of $t$ (days after the merging) for
 selected trajectories (top-left) and those mass-averaged (top-right;
 also shown are those from $\beta$-decay, fission, and
 $\alpha$-decay). In each panel, the heating rates for the solar
 $r$-process pattern ($\dot{q}_{\mathrm{solar-}r}$) and the anaylytical
 approximation ($\dot{q}_\mathrm{analytic}$) are shown by black-solid
 and short-dashed lines, respectively. Lower panels are the same as the
 upper panels but for those relative to
 $\dot{q}_\mathrm{analytic}$. Long-dashed lines indicate the factor of 2
 ranges from unity (short-dashed line).}

\label{fig:qdot}
\end{figure*}

The $r$-processing ends a few 100~ms after the merging. The subsequent
abundance changes by $\beta$-decay, fission, and $\alpha$-decay are
followed up to $t = 100$~days; the resulting radioactive heating is
relevant for kilonova emission. Figure~\ref{fig:qdot} displays the
temporal evolutions of the heating rates for selected trajectories
(top-left) and those mass-averaged (top-right). For comparison purposes,
the heating rate for the nuclear abundances with the solar $r$-process
pattern \citep[for $A \ge 90$, $\dot{q}_{\mathrm{solar-}r}$; same as
that used in][]{Hotokezaka2013a, Tanaka2014}, $\beta$-decaying back from
the neutron-rich region, is also shown in each panel. The short-dashed
line indicates an analytical approximation defined by
$\dot{q}_\mathrm{analytic} \equiv 2 \times 10^{10}\, t^{-1.3}$ \cite[in
units of erg~g$^{-1}$~s$^{-1}$; $t$ is time in day,
e.g.,][]{Metzger2010}. Lower panels show the heating rates relative to
$\dot{q}_\mathrm{analytic}$.

Overall, each curve reasonably follows $\dot{q}_\mathrm{analytic}$ by
$\sim 1$~day. After this time, the heating is dominated by a few
radioactivities and becomes highly dependent on
$Y_\mathrm{e}$. Contributions from the ejecta of $Y_\mathrm{e} > 0.3$
are generally unimportant after $\sim 1$~day. We find that the heating
for $Y_\mathrm{e} = 0.34$ turns to be significant after a few 10 days
because of the $\beta$-decays from $^{85}$Kr (half-life of $T_{1/2} =
10.8$~yr; see Figure~\ref{fig:ymav} for its large abundance), $^{89}$Sr
($T_{1/2} = 50.5$~d), and $^{103}$Ru ($T_{1/2} = 39.2$~d). Heating rates
for $Y_\mathrm{e} = 0.19$ and 0.24, whose abundances are dominated by
the second peak nuclei, are found to be in good agreement with
$\dot{q}_{\mathrm{solar-}r}$. This is due to a predominance of
$\beta$-decay heating from the second peak abundances, e.g., $^{123}$Sn
($T_{1/2} = 129$~d) and $^{125}$Sn ($T_{1/2} = 9.64$~d) around a few
10~days.

Our result shows that the heating rate for the lowest $Y_\mathrm{e}$ (
$= 0.09$) is the greatest after 1~day with a few times larger values
than those in previous works \cite[with $Y_\mathrm{e} \sim 0.02$--0.04
in][]{Goriely2011, Rosswog2014}. In our case, the radioactive heating is
dominated by the spontaneous fissions of $^{254}$Cf and $^{259,
262}$Fm. It should be noted that the heating from spontaneous fission is
highly uncertain because of the many unknown half-lives and decay modes
of nuclides reaching to this quasi-stable region ($A \sim 250$--260 with
$T_{1/2}$ of days to years). In fact, tests with another set of
theoretical estimates show a few times smaller rates after $\sim 1$~day
(because of diminishing contributions from $^{259, 262}$Fm), being
similar to the previous works. It appears difficult to obtain reliable
heating rates with currently available nuclear data when fission plays a
dominant role.

In our result the total heating rate is dominated by $\beta$-decays all
the times because of the small ejecta amount of $Y_\mathrm{e} <
0.15$. The radioactive heating after $\sim 1$~day is mostly due to the
$\beta$-decays from a small number of species with precisely measured
half-lives. Uncertainties in nuclear data are thus irrelevant. The
mass-averaged heating rate for $t \sim 1$--10~days is smaller than
$\dot{q}_\mathrm{analytic}$ and $\dot{q}_{\mathrm{solar-}r}$ because of
the overabundances near $A = 100$ (Figure~\ref{fig:ymav}, bottom) that
do not significantly contribute to heating. The differences are,
however, well within about a factor of two. In conclusion, if merger
ejecta have a solar $r$-process-like abundance pattern,
$\dot{q}_{\mathrm{solar-}r}$ (and $\dot{q}_\mathrm{analytic}$) serves as
a good approximation for kilonova emission\footnote{These heating rates
correspond to the heating efficiency, defined by $f \equiv \dot{Q}\,
t_\mathrm{peak} / M_\mathrm{ej}\, c^2$ \citep[$\dot{Q}$ and
$t_\mathrm{peak}$ are the total heating rate and peak time of a kilonova
transient,][]{Li1998}, of $f/10^{-6} \approx 1$ and 0.5 for
$t_\mathrm{peak} = 1$ and 10~days, respectively, with the thermalization
factor of 0.5 \citep{Metzger2010}.}.

It is important to note that our merger simulation exhibits different
$Y_\mathrm{e}$ distributions between the orbital and polar directions
(Figure~\ref{fig:histogram}). Multi-dimensional information of nucleosynthetic
abundances will be needed when we discuss the angler dependences of
kilonova emission \citep{Roberts2011, Grossman2014}.

\section{Summary}\label{sec:summary}

We examined $r$-process calculations based on the full-GR, approximate
neutrino transport simulation of the NS-NS merger with the equal masses
($= 1.3\, M_\odot$) of NSs. Different from previous studies, the merger
ejecta exhibited a wide range of $Y_\mathrm{e} \approx 0.09$--0.45 that
led to the nucleosynthetic abundance distribution being in good
agreement with the solar $r$-process pattern. Given that the model is
representative, our result (with the present estimate of the Galactic
event rate) implies that NS-NS mergers can be the major origin of all
the $r$-process elements in the Galaxy.

Our result also indicates that the radioactive heating (that powers a
kilonova transient) after $\sim 1$~day from the merging is dominated by
the $\beta$-decays of a small number of species with measured
half-lives. The total heating rates are thus well approximated by the
$\beta$-decays of the solar $r$-process-like abundances as well as by
the approximation of $\propto t^{-1.3}$. Detailed multi-dimensional
information of nucleosynthesis abundances should be, however, taken into
account when we consider the spatial dependences of kilonova emission.

Our result implies that the previous thought of NS-NS merger events,
dynamically ejecting almost pure NS matter, should be reconsidered. The
shock-heated and neutrino-processed ejecta from a HMNS are in fact
modestly neutron-rich: the phenomenon similar to the early stage of a
CCSN (a proto-NS instead of a HMNS). Much more works will be needed to
test if similar results are obtained with full-3D nucleosynthetic
analyses, with different NS masses and their ratios, with other
(reasonable) EOSs, with higher spatial resolution, etc. Nucleosynthetic
contributions from BH-NS mergers, as well as from the BH-accretion tori
subsequent to NS-NS/BH-NS mergers, should be also explored to draw
conclusions on the role of CBMs to the Galactic chemical evolution of
the $r$-process nuclei.

\acknowledgements

We are grateful to S. Goriely and T. Tachibana for providing the data of
fission properties and $\beta$-decay rates and M. Hempel for the EOS
table. The project was supported by the RIKEN iTHES Project, the JSPS
Grants-in-Aid for Scientific Research (23740160, 24244028, 24740163,
25103510, 25103512, 25105508, 26400232, 26400237), Grant-in-Aid for
Scientific Research on Innovative Area (20105004), and
EU-FP7-ERC-2012-St Grant 306901. Koutarou Kyutoku is supported by JSPS
Postdoctoral Fellowships for Research Abroad. CBM simulations were in
part performed on Cray XC30 at CfCA of NAOJ and Fujitsu FX10 at
(Information Technology Center of) the University of Tokyo. This work
was in part developed during the long-term workshop on Gravitational
Waves and Numerical Relativity held at the Yukawa Institute for
Theoretical Physics, Kyoto University in May and June 2013.

\end{document}